\documentclass[preprint2]{aastex}

\usepackage{CJK}
\usepackage{longtable}
\usepackage{lipsum} 
\usepackage{multirow}
\usepackage{natbib}
\usepackage{epsfig}
\usepackage{graphicx}
\usepackage{bm}
\usepackage{amsmath}
\usepackage{amsfonts}
\usepackage{amssymb}
\usepackage{xcolor, soul}
\usepackage{mathrsfs}
\usepackage[caption=false]{subfig}
\usepackage{ulem}
\usepackage{hyperref}
\usepackage{orcidlink}

\def\arcsec{$^{\prime\prime}$}

\def\deg{$^{\circ}$}

\graphicspath{{./}{figures/}}

\begin{document}

\title{Submillimeter Observations of Magnetic Fields in Massive Star-forming Region W75N}

\author{Lingzhen Zeng\orcidlink{0000-0001-6924-9072}\altaffilmark{1}, Qizhou Zhang\orcidlink{0000-0003-2384-6589}\altaffilmark{1}, Felipe O. Alves\orcidlink{0000-0002-7945-064X}\altaffilmark{2}, Tao-Chung Ching\orcidlink{0000-0001-8516-2532}\altaffilmark{3,4}, Josep M. Girart\orcidlink{0000-0002-3829-5591}\altaffilmark{5,6} and Junhao Liu\orcidlink{0000-0002-4774-2998}\altaffilmark{7}}

\altaffiltext{1}{Center for Astrophysics $|$ Harvard $\&$ Smithsonian, 60 Garden Street, Cambridge, MA 02138, USA}
\altaffiltext{2}{Center for Astrochemical Studies, Max-Planck-Institut f\"{u}r extraterrestrische Physik (MPE), Gie$\beta$enbachstr. 1, D-85741 Garching, Germany}
\altaffiltext{3}{Research Center for Intelligent Computing Platforms, Zhejiang Lab, Hangzhou 311100, People’s Republic of China}
\altaffiltext{4}{National Radio Astronomy Observatory, 1003 Lopezville Road, Socorro, NM 87801, USA}
\altaffiltext{5}{Institut de Ci\`{e}ncies de l’Espai (ICE), CSIC, Can Magrans s/n, E-08193 Cerdanyola del Vall\`{e}s, Catalonia, Spain}
\altaffiltext{6}{Institut d’Estudis Espacials de Catalunya (IEEC), E-08034 Barcelona, Catalonia, Spain}
\altaffiltext{7}{East Asian Observatory, 660 N. A’oh\={o}k\={u} Place, University Park, Hilo, HI 96720, USA}








\begin{abstract}
  This paper presents the results of full polarization observations of the massive star-forming region W75N, conducted with $\sim$3\arcsec$\;$spatial resolutions at 345 GHz using the Submillimeter Array (SMA). The magnetic field structures in the dense cores of the region are derived using the linearly polarized continuum emission. The overall magnetic field strength and orientation are found to agree with those from the previous observations. The plane-of-sky (POS) component of the magnetic field in the region was calculated to be $\sim$0.8 $\pm$ 0.1 mG using the angular dispersion function (ADF) method. Further analyses involving the polarization-intensity gradient-local gravity method and H$^{13}$CO$^+$ (4--3) line data indicated that the cloud is undergoing global gravitational collapse and the magnetic field is shaped by gravity and outflows in the dense core regions.

\end{abstract}



\section{Introduction} 
\label{sec:intro}

Stars are born in dense molecular cores when self gravity exceeds the internal support and drives gravitational collapse and the formation of an embedded protostar. In addition to gravity, both turbulence and magnetic fields influence the dynamical evolution of the molecular gas and impact the outcome of star formation. Solenoidal turbulence suppresses star formation since it acts similar to the thermal pressure that counteracts gravity, thus it hinders star formation. Compressive turbulence, on the other hand, compresses the gas and enhances its densities, thus promotes star formation \citep{2020MNRAS.493.4643M, 2010A&A...512A..81F}. Magnetic fields, well coupled with the molecular gas, tend to restrict the movement of material along the field lines, thus hinder the star formation \citep{2009ApJ...702.1428H, 2014MNRAS.439.3420M}.

There have been considerable efforts devoted to accessing the role of magnetic fields in star forming dense molecular cores. Thanks to the improvement in sensitivity, polarimetric observations in the millimeter and sub-millimeter wavelengths become increasingly accessible to probe the plane-of-the-sky component of magnetic fields through linearly polarized dust emission \citep{2014ApJ...792..116Z}. We refer readers to recent reviews on the development of observational efforts on magnetic fields in molecular clouds and star formation \citep{2019FrASS...6....3H, 2019FrASS...6...15P, 2022FrASS...9.3556L, 2022arXiv220311179P}. 


Despite the considerable progress, there is a lack of understanding on how magnetic fields may affect star formation in a protocluster environment where multiple stars arise from collapse and fragmentation of molecular gas. We present Submillimeter Array observations of W75N, a massive star-forming region that contains a number of \ion{H}{2} regions and is located in the local spiral arm at a distance of approximately 1.3 kpc \citep{2012A&A...539A..79R}. W75N is part of the Cygnus-X giant molecular cloud, which spans over 100 pc and includes the renowned DR21 region.

Early observations of the region indicated that W75N IRS 1, a cluster of young stellar objects (YSOs), powered a massive molecular outflow. VLA observations at 4.9 GHz detected three ionized regions, W75N(A), W75N(B), and W75N(C) near the center of the outflow \citep{1981ApJ...244...76H}. Later, VLA 8.4 GHz observations revealed that W75N(B) consisted of three compact regions, Ba, Bb, and Bc \citep{1994A&A...284..215H}. Using 1.3 cm continuum VLA observations, \cite{1997ApJ...489..744T} discovered VLA 1 (Ba), VLA 3 (Bb), and another compact source located between them (VLA 2). \cite{2010AJ....139.2433C} suggested that source Bc was a radio Herbig–Haro object \citep{2000prpl.conf..815E} powered by the VLA 3 radio jet. They also discovered the VLA 4 source, located south of the VLA 1–VLA 3 group. \cite{2013ApJ...767...86K} found that the outflow of VLA 2 was in a transition from a shell-like to a more elongated jet-like shape based on VLBI observations of 22 GHz water masers. Further observations by \citep{2014A&A...565L...8S} showed that the water maser distribution around VLA 1 was stable, while the shell-like structure in VLA 2 was expanding along the direction parallel to the thermal radio jet of VLA 1, which was later confirmed by \cite{2015Sci...348..114C}. Recently, using VLA-A data covering 4 to 48 GHz, \cite{2020MNRAS.496.3128R} concluded that Bc and VLA 4 were obscured Herbig–Haro objects excited by the jet from VLA 3.

Observations in millimeter wavelengths have revealed the presence of 9 dense cores (MM1 to MM9) in the W75N region. These were identified using continuum data obtained with BIMA and CARMA \citep{2001ApJ...546..345S, 2002ApJ...577..260W, 2003ApJ...584..882S, 2004ApJ...601..952S}. The MM1 core was further studied using the SMA and resolved into two compact continuum sources, MM1a and MM1b \citep{2010ApJ...723.1231M}. In addition, a dense core labeled as MM[N] was recently reported to the north of MM1 using ALMA data at 1.3 mm \citep{2020MNRAS.496.3128R}.

Previous polarization observations of W75N at 450, 870, and 1100 $\mu$m, using the JCMT, yielded only a single polarization segment due to the large beamsizes of around 12\arcsec--19\arcsec\, \citep{1995A&A...294..831V, 1998A&A...333L..23G}. At 870 and 1100 $\mu$m, the inferred magnetic field had an average position angle of approximately 150\deg, while at 450 $\mu$m, it was measured to be around 37\deg. To improve the angular resolution, we conducted full polarization observations of the W75N region using the SMA with spatial resolutions of approximately 3\arcsec\, at 345 GHz. In this study, we focus on the central region of W75N, which includes the MM1 to MM4, and MM[N] cores. We present the derived parameters of these dense cores using the dust continuum polarization data in this paper. We summarize the SMA observations in Section~\ref{sec:obs} and present the results in Section~\ref{sec:res}. A discussion of the results is shown in Section~\ref{sec:disc}, followed by a summary in Section~\ref{sec:sum}.

\section{Observations and data reduction} 
\label{sec:obs}
The observations of W75N were carried out between 2012 July 03 and 2012 Aug 09 with the Submillimeter Array (SMA) \citep{2004ApJ...616L...1H}. Three observations were made in July using the compact array configuration, and three were made in August using the subcompact configuration. The number of antennas in the array varied between 6 and 7. The observational parameters and calibration sources can be found in Table~\ref{tab:obs_sum}. The Application Specific Integrated Circuit (ASIC) correlator provided a 4 GHz IF bandwidth (4-8 GHz) with a uniform spectral width of 812.5 kHz per channel. The receivers were tuned to the 345 GHz band, which captured the CO (3-2) and H$^{13}$CO$^+$ (4--3) lines, with a velocity resolution of approximately 0.70 km s$^{-1}$.

The visibility data from the observations were calibrated for bandpass, flux, and time-dependent gains using the IDL superset MIR package adapted for the SMA \citep{1993PASP..105.1482S}. The calibrated data were then exported to the Miriad \citep{1995ASPC...77..433S} format for instrumental polarization calibrations and imaging. Table~\ref{tab:obs_sum} lists the calibrators used for each track. The synthesized beam size of the combined visibilities was approximately 3.05\arcsec $\times$ 2.83\arcsec. The 1$\sigma$ RMS noise of the Stokes I image of the continuum emission was approximately 26.1 mJy~beam$^{-1}$, while the RMS noise of the Stokes Q/U maps after de-biasing using the method from \cite{2006PASP..118.1340V} was approximately 1.4 mJy~beam$^{-1}$. The Astropy package \citep{astropy:2013, astropy:2018} was used for the final analysis.

\begin{deluxetable}{ccccccc}
\tablecaption{Observational Summary}
\label{tab:obs_sum}
\tablewidth{0pt}
\tablehead{
\colhead{Observation} & Number of &\colhead{Array} & \colhead{Baseline} & \colhead{Flux} & \colhead{Gain} & \colhead{Polarization and }\\
\colhead{date} & Antennas &\colhead{configuration} &\colhead{range (m)} & \colhead{calibrator} & \colhead{ calibrator} & \colhead{bandpass calibrator}}
\startdata
2012 Jul 03 & 7 & compact & 16.5 -- 32 &  Titan, Uranus & MWC349A & 3c279 \\
2012 Jul 04 & 7 & compact & 16.5 -- 32 & Titan & MWC349A & 3c279 \\ 
2012 Jul 05 & 7 & compact & 16.5 -- 32 & Titan, Uranus & MWC349A & 3c84 \\ 
2012 Aug 07 & 6 & subcompact & 9.5 --25 & Uranus & MWC349A & 3c84 \\ 
2012 Aug 08 & 6 & subcompact & 9.5 -- 25 & Uranus & MWC349A & 3c84 \\ 
2012 Aug 09 & 6 & subcompact & 9.5 -- 25 & Uranus & MWC349A & 3c84 \\
\enddata
\end{deluxetable}

\section{Results} 
\label{sec:res}

\subsection{Continuum emission} 
\label{subsec:con}
\begin{figure}[ht!]
\epsscale{2.0}
\plottwo{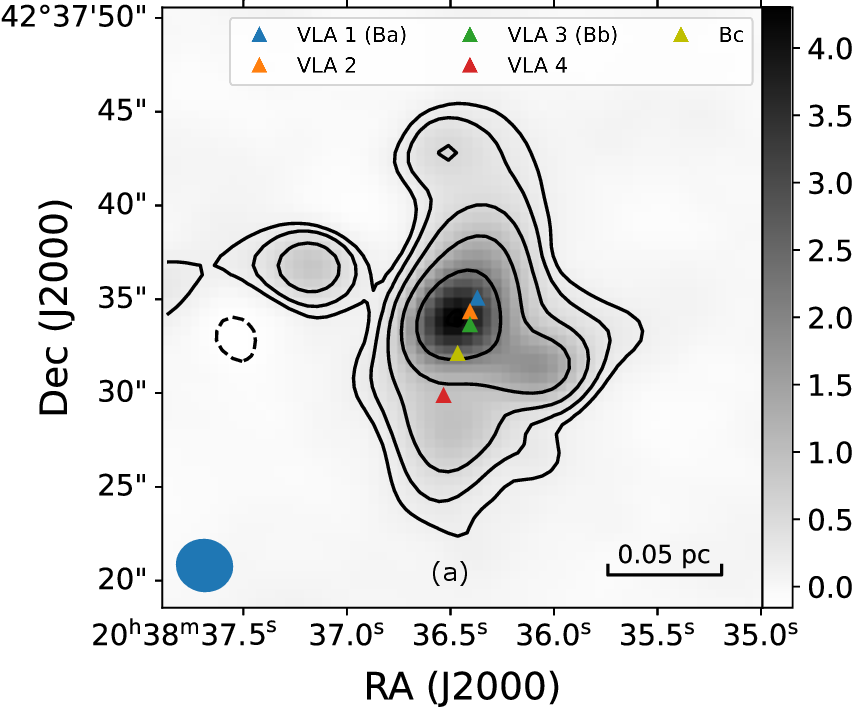}{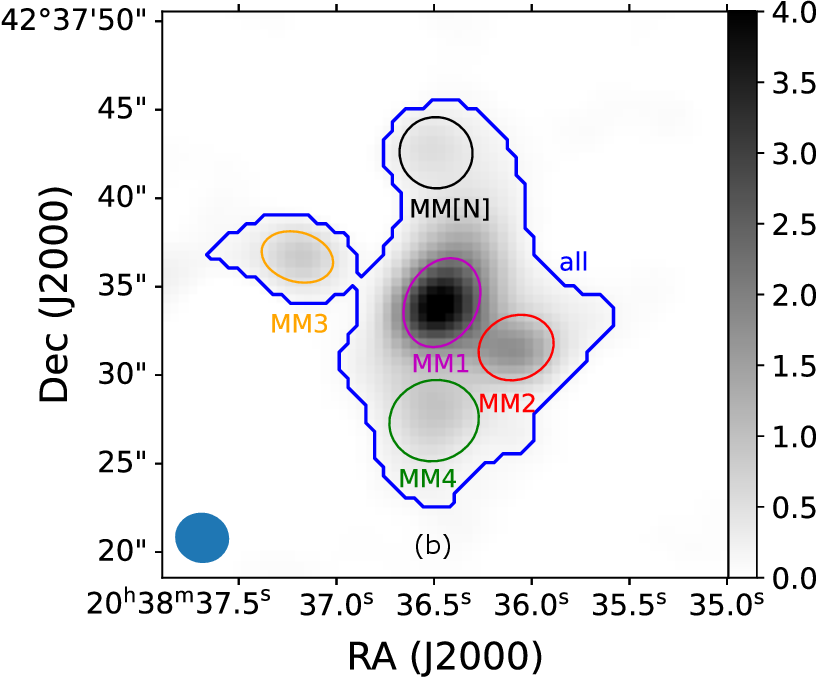}
\caption{(a) 345 GHz continuum emission image of W75N. The synthesized beam marked in the lower-left corner of the image is 3.05\arcsec $\times$ 2.83\arcsec $\;$at P.A. = 77.8\deg. The intensity is shown in gray scales in units of Jy~beam$^{-1}$ with contours levels at [$-$5, 5, 10, 20, 40, 80, 160, and 320 $\sigma$], where $\sigma$ = 26.1 mJy~beam$^{-1}$. The previous detected compact sources are denoted by filled triangles. (b) Dense cores identified by CARTA are labeled as MM1, MM2, MM3, MM4 and MM[N]. The ellipses show the FWHMs from 2D Gaussian fits, while the blue color mask of the entire region is labeled as ``all''.}
\label{fig:i}
\end{figure}

Figure~\ref{fig:i} (a) illustrates the 345 GHz continuum emission of the W75N region. To identify dense structures in this area, we applied the dendrogram algorithm \citep{2008ApJ...679.1338R} to the continuum data using the astrodendro\footnote{\url{http://www.dendrograms.org/}} package. For the astrodendro analysis, we set the minimum value for the structure to be considered as 3$\sigma$, the minimum height required for an independent structure to be retained as 1$\sigma$, and the minimum number of pixels for a structure to be considered as half of the synthesized beam area. Using the astrodendro results as the initial input, we performed a final 2D Gaussian fit to each of the identified cores using the Cube Analysis and Rendering Tool for Astronomy (CARTA) \citep{2021ascl.soft03031C}. We followed the nomenclature for dense cores used in \citet{2003ApJ...584..882S} and \citet{2020MNRAS.496.3128R}. The mask for the entire cloud and the FWHM ellipses representing the dense cores are shown in Figure~\ref{fig:i} (b). Table~\ref{tab:dense_str_obs} lists the observation parameters for those structures. The paremeters for the ``all'' mask are from astrodendro, and the equivalent FWHMs are calculated from the intensity weighted second moment in the corresponding directions. The parameters of the dense cores are from CARTA. 

\begin{deluxetable}{ccccccc}
\tablecaption{Observation parameters of dense structures}
\label{tab:dense_str_obs}
\tablewidth{0pt}
\tablehead{
\multirow{2}{*}{Structure} & \colhead{RA(J2000)} & \colhead{Dec(J2000)} & \colhead{Inegrated Flux} & \colhead{FWHM} & \colhead{Peak Intensity} & \colhead{PA}\\
 & \colhead{(h:m:s)} & \colhead{(d:m:s)} & \colhead{($\mathrm{Jy}$)} & \colhead{a \arcsec $\;\times$ b \arcsec} & \colhead{($\mathrm{Jy}$ beam$^{-1}$)} & \colhead{(\deg)}
}
\startdata
all & 20 38 36.44 &	42 37 33.75 & 17.9 & 9.2 $\times$ 7.4 & 2.3 & 19.6\\
MM1 & 20 38 36.46 &	42 37 34.11 & 10.2 & 5.2 $\times$ 4.1 & 4.1 & 146.9\\
MM2	& 20 38 36.08 &	42 37 31.54 & 3.0 & 4.5 $\times$ 3.5 & 1.7 & 106.1\\
MM3	& 20 38 37.20 &	42 37 36.69 & 1.1 & 4.1 $\times$ 2.8 & 0.9 & 75.2\\
MM4	& 20 38 36.50 &	42 37 27.57 & 2.3 & 5.2 $\times$ 4.6 & 0.8 & 98.7\\
MM[N] & 20 38 36.49 & 42 37 42.57 & 1.1 & 4.1 $\times$ 4.0 & 0.5 & 73.2\\
\enddata
\end{deluxetable}

\begin{deluxetable}{cccccc}
\tablecaption{Fitting parameters of dense structures}
\label{tab:dense_str_fit}
\tablewidth{0pt}
\tablehead{
\multirow{2}{*}{Structure} & \colhead{$T_d$} & \colhead{$M$} & \colhead{$\rho$} & \colhead{N$_{\text{H}_{2}}$} & \colhead{n$_{\text{H}_{2}}$}\\
 & \colhead{(K)} & \colhead{(M$_\odot$)} & \colhead{(10$^{-15}$ kg m$^{-3}$)} & \colhead{(10$^{23}$ cm$^{-2}$)} & \colhead{(10$^{6}$ cm$^{-3}$)}
}
\startdata
all & 63 & 35.5 & 4.1 & 1.8 & 0.85\\
MM1 & 73 & 17.1 & 11.2 & 2.8 & 2.3\\
MM2	& $<$45 & 8.9 & 9.2 & 2.0 & 1.9\\
MM3	& $<$45 & 3.4 & 5.6 & 1.0 & 1.2\\
MM4	& 58 & 5.0 & 2.8 & 0.73 & 0.58\\
MM[N] &45 & 3.1 & 3.0 & 0.67 & 0.63\\
\enddata
\end{deluxetable}

Assuming that the cloud is isothermal, the continuum emission is optically thin, and the gas-to-dust mass ratio is a constant $\Lambda = 100$, we can derive the the total mass of the structures using the observed integrated flux of the dust emission, F$_{\nu}$, by
\begin{equation}
\label{eq:m}
M=\dfrac{\Lambda \text{F}_{\nu} D^2}{B_{\nu}(T_d)\kappa_{\nu}},
\end{equation}
where $D$ = 1.3 kpc is the distance to the source, $\kappa_{\nu} = (\nu/1000 \text{GHz})^{\beta}$ m$^2$ kg$^{-1}$ is the dust opacity \citep{1983QJRAS..24..267H}, and $B_{\nu}(T_d)$ is the Planck function at a given dust temperature $T_d$. We utilized an opacity index of $\beta = 1.5$ \citep{1994ApJ...421..615P, 2007A&A...466.1065B, 2007ApJ...654L..87C}, and the average $T_d$ within each dense structure in W75N was listed in Table~\ref{tab:dense_str_fit} from ammonia hyper-fine line fitting using EVLA data (Zhang, X., et al. 2023, in prep.). While the fittings for MM2 and MM3 did not converge, we were still able to estimate the temperatures to be between 30 K and 45 K, and hence, we used $T_d$ = 45 K to determine the lower limits for the mass. The average density, column density N$_{\text{H}_{2}}$ and volume density n$_{\text{H}_{2}}$ within each structure are calculated as:
\begin{equation}
\label{eq:rho}
\rho = \frac{3M}{4\pi r^3},
\end{equation}
\begin{equation}
\label{eq:cd}
\text{N}_{\text{H}_{2}} = \frac{M}{\pi m_{\text{H}}\mu_{\text{H}_2} r^2},
\end{equation}
\begin{equation}
\label{eq:vd}
\text{n}_{\text{H}_{2}} = \frac{3M}{4\pi m_{\text{H}}\mu_{\text{H}_2} r^3},
\end{equation}
where $r = \sqrt{(\text{FWHM}_a  \times \text{FWHM}_b)}$ \citep{2007A&A...476.1243M} is the geometric mean radius of the structure,  $\mu_{\text{H}_2}$ = 2.86 is the mean hydrogen molecular weight \citep{2013MNRAS.432.1424K, 2015MNRAS.450.1094P}, and $m_{\text{H}}$ is the hydrogen atomic mass. The mass, average density, column density, and volume density of the dense structures derived from Equations~\eqref{eq:m} to \eqref{eq:vd} are listed in Table~\ref{tab:dense_str_fit}. The estimated column and volume densities of the structures in W75N are generally similar to those in other massive star-forming regions. 

The uncertainties in the parameters discussed above arise from various sources. The characterization of the constant $\Lambda$ and $\kappa_\nu$ is not well-constrained and contributes to an uncertainty over 50\% \citep{Draine2011, 2018A&A...617A.100B} and a factor of 2 \citep{1994A&A...291..943O, 2007A&A...476.1243M}, respectively. The ammonia line data yields dust temperatures ranging from 30 -- 73 K, consistent with the results of \cite{2001ApJ...546..345S}, which estimated temperatures of 35--75 K. For MM2 and MM3, we estimated the lower mass limits using the upper fitting temperatures. The distance to W75N, as estimated by \cite{2012A&A...539A..79R}, is uncertain by approximately 5\%. As a result, the final uncertainties for the mass, density, column density, and volume density listed in Table~\ref{tab:dense_str_fit} are estimated to be at least a factor of 2.1.
 
\subsection{Dust polarization} 
\label{subsec:pol}
Since polarized intensity and polarized percentage are defined as positive values, the measurements of these two parameters tend to be biased towards larger values. In order to correct for this bias, the debiased polarized intensity (PI) can be calculated using the following formula \citep{2006PASP..118.1340V}:

\begin{equation}
\label{eq:pi}
\text{PI} = \sqrt{Q^2+U^2-0.5(\sigma^2_Q + \sigma^2_U)},
\end{equation}
where $\sigma_Q$ and $\sigma_U$ are the 1$\sigma$ rms noise of the $Q$ and $U$ maps. The polarization fraction is calculated as:

\begin{equation}
\label{eq:pf}
\text{Pf} = \text{PI}/\text{I},
\end{equation}
where I is the Stokes I intensity. 

\begin{figure}[ht!]
\epsscale{1}
\plotone{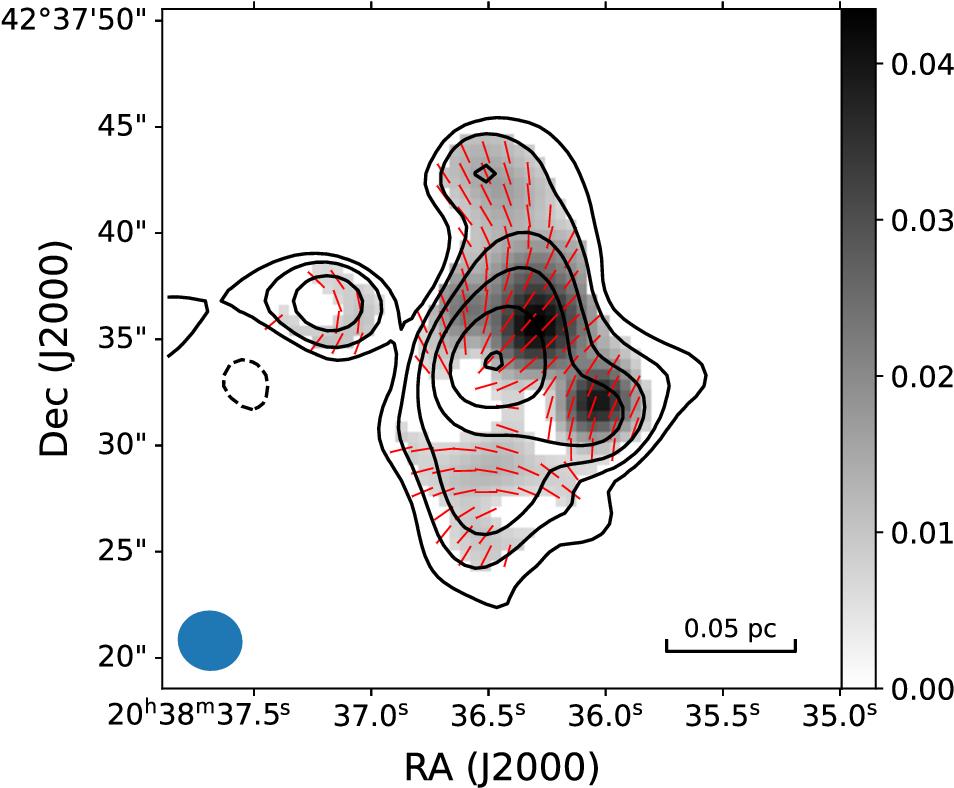}
\caption{W75N magnetic field orientation map. Polarized emission intensities are shown in gray scales in the units of Jy~beam$^{-1}$. Magnetic field orientations inferred from the linearly polarized emission with PI/$\sigma_{PI}$ $>$ 3 are shown as red line segments. Intensity contour levels are the same as those in Figure~\ref{fig:i}.}
\label{fig:polvec}
\end{figure}

\begin{figure}[ht!]
\epsscale{1}
\plotone{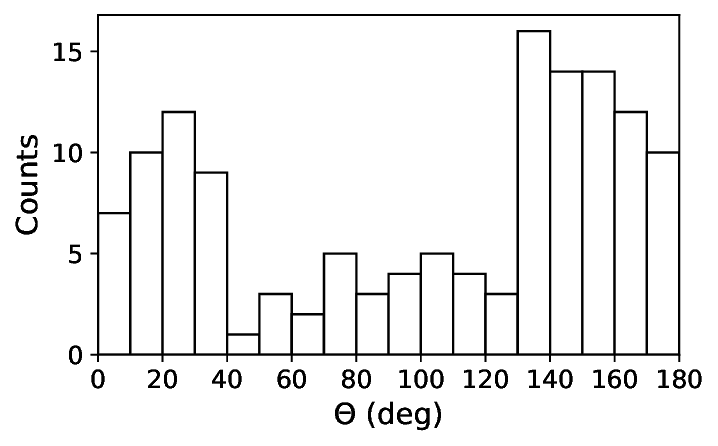}
\caption{Histogram of position angles of the magnetic field segments from Figure~\ref{fig:polvec}. The bin size is 10\deg, and position angles are measured east of north.}  
\label{fig:pv_hist}
\end{figure}

Assuming that irregular grains have their shortest axis aligned with the magnetic field lines \citep{2007JQSRT.106..225L, 2007MNRAS.378..910L, 2015ARA&A..53..501A}, we can determine the magnetic field orientation projected on the plane of sky (POS) by rotating the polarization segments by 90\deg. Figure~\ref{fig:polvec} displays the magnetic field orientations overlaid on the polarization intensity map, where two polarization intensity peaks are observed, one close to MM2 and the other to the northwest of MM1. As shown in Figure~\ref{fig:pv_hist}, the magnetic field orientation distribution falls into three major groups. The first group with position angles between 0\deg\, and 40\deg\, is dominated by the polarized emission from MM[N], while the second group with position angles between 60\deg\, and 120\deg\, is mainly associated with the polarized emission from MM4. The last group comprises detections from the polarization intensity peaks around MM1 and MM2, with polarization angles from 130\deg\, to 180\deg. As these groups are found to be related to the dense structures described in Section~\ref{subsec:con}, the magnetic field angles can be assumed to be uniform within each dense structure. 

\begin{figure}[ht!]
\epsscale{1}
\plotone{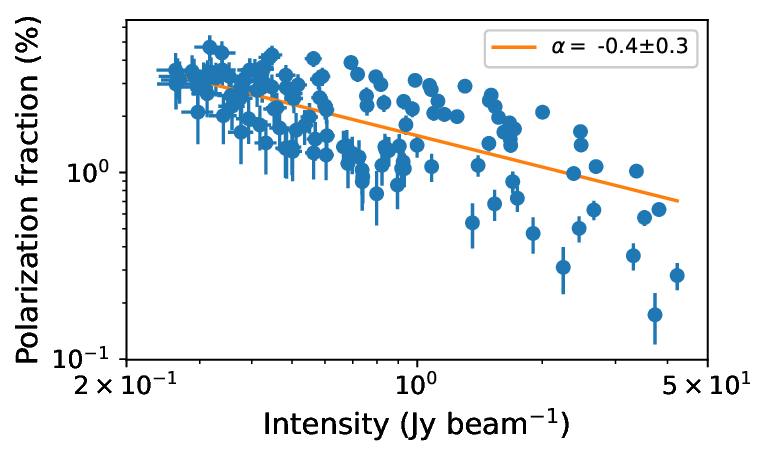}
\caption{Polarization percentage vs. Stokes I plots. Polarization data points from Figure~\ref{fig:polvec} are shown in log-log scale with error bars. The orange line shows the fitting results using the power-law model.}
\label{fig:pf_i}
\end{figure}

In Figure~\ref{fig:pf_i}, we present the polarization fraction (Pf) as a function of I for the entire W75N region. We then fitted the Pf--I relation using a simple power law of P $\propto$ I$^{\alpha}$, with an estimated index of $\alpha = -0.4 \pm 0.3$. This relation can be used to evaluate the grain alignment efficiency within a cloud. In more developed star forming regions, the alignment efficiency is often enhanced by additional radiation, resulting in a power law index with smaller absolute value (the slope is shallower) in the Pf--I relation.

\subsection{Magnetic field analysis}
\label{subsec:ADF}

The Davis-Chandrasekhar-Fermi (DCF) method \citep{1951PhRv...81..890D, 1953ApJ...118..113C} relates the dispersion of polarization position angles to the large scale mean magnetic field strength. This analysis tool has been widely used to obtain the strength of the magnetic field projected on the POS. We refer readers to \citep{2022arXiv220806492L} for a review and detailed discussion of the assumptions in the DCF analysis. Further studies have been made to expand the DCF method using the angular dispersion function (ADF) analysis \citep{2008ApJ...679..537F, 2009ApJ...696..567H, 2009ApJ...706.1504H, 2011ApJ...733..109H, 2016ApJ...820...38H}. Specifically, using the twin Gaussian model for the interferometer beams, \cite{2016ApJ...820...38H} derived the angular dispersion solutions for the interferometer, which can be expressed as equation (13) in their work. We can rewrite it as: 

\begin{equation}
\label{eq:adf}
\begin{split}
1-\left<\text{cos}[\Delta\phi(\ell)]\right> &= \sum_{j=1}^{\infty}a_{2j}\ell^{2j} + \frac{1}{1+N\left<B_0^2\right>/\left<B_t^2\right>} \\
&\quad -b^2(\ell),
\end{split}
\end{equation}
where $\Delta\phi(\ell)$ is the angular difference of the two polarization segments separated by a distance of $\ell$, $N$ is the number of the turbulent cells, $\left<B_0^2\right>/\left<B_t^2\right>$ is the large scale to turbulent magnetic strength ratio, and $b^2(\ell)$ is the local turbulent component of the angular dispersion function. The contribution of the large scale component to the dispersion function can be written as $1-\left<\text{cos}[\Delta\phi(\ell)]\right> - b^2(\ell)$. Assuming the turbulent correlation length is $\delta$, the effective thickness of the observation region is $\Delta^\prime$, and the beamsizes (standard deviation) of the twin Gaussian model are $W_1$ and $W_2$, $N$ and $b^2(\ell)$ can be written as:
\begin{align}
\label{eq:N}
&N_1 = \frac{(\delta^2+2W_1^2)\Delta^\prime}{\sqrt{2\pi}\delta^3}\\
&N_2 = \frac{(\delta^2+2W_2^2)\Delta^\prime}{\sqrt{2\pi}\delta^3}\\
&N_{12} = \frac{(\delta^2+W_1^2+W_2^2)\Delta^\prime}{\sqrt{2\pi}\delta^3}\\
&N = \left(1/N_1+1/N_2-2/N_{12}\right)^{-1}
\end{align}

\begin{equation}
\begin{split}
\label{eq:turb}
b^2(\ell) &= \frac{N}{1+N\left<B_0^2\right>/\left<B_t^2\right>}\bigl\{\frac{1}{N_1}e^{-\ell^2/[2(\delta^2+2W_1^2)]} \\
& \quad +\frac{1}{N_2}e^{-\ell^2/[2(\delta^2+2W_2^2)]}\\
& \quad -\frac{1}{N_{12}}e^{-\ell^2/[2(\delta^2+W_1^2+W_2^2)]}\bigr\}.
\end{split}
\end{equation}

\begin{figure}[ht!]
\epsscale{1}
\plotone{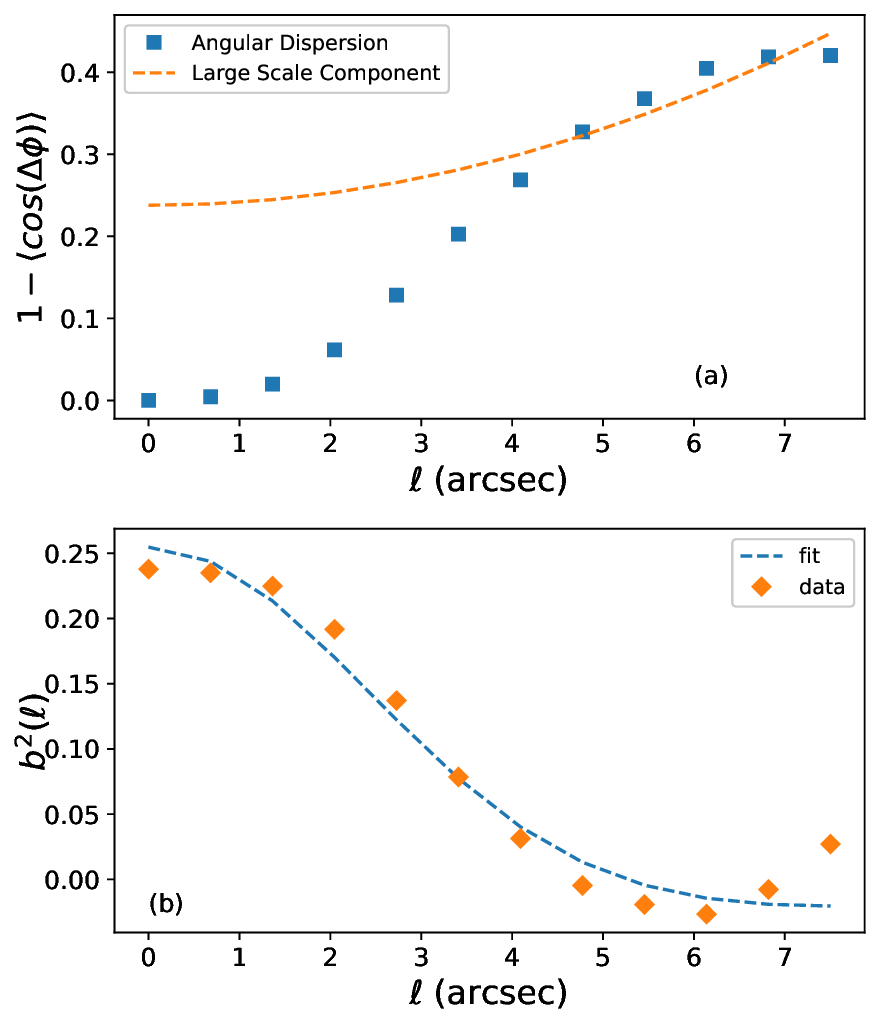}
\caption{Dispersion function 1 - $\left<\text{cos}(\Delta\phi)\right>$ for W75N using the 345 GHz SMA data. (a) Angular dispersion data (symbols) plotted as a function of distance $\ell$, the dashed curve from Gaussian fitting represent contribution from the large scale component. (b) shows the resulting turbulence correlation function $b^2(\ell)$ and its fitting curve.} 
  \label{fig:ADF}
\end{figure}

Due to the limited number of detected polarization segments, performing the angular dispersion analysis on each dense structure in the W75N region is impractical. Therefore, we estimated the mean magnetic field ($B_0$) for the entire cloud by utilizing the position angle data from the polarization measurements presented in Figure~\ref{fig:polvec}. For the twin Gaussian beamsizes, the telescope beamwidth radius $W_1$ can be estimated using the size of the synthesized beam, $W_1 = \sqrt{\text{FWHM}_{a}^{\text{beam}}\times \text{FWHM}_b^{\text{beam}}}/(\sqrt{8\text{ln}2})$, and $W_2$ is the resolution calculated from the shortest baseline of the array. For our analysis, we set $W_1$ = 1.2\arcsec\, and $W_2$ = 8.0\arcsec. We determined the effective thickness of the cloud to be the ratio of the volume to the cross area of the equivalent sphere of the entire cloud:
\begin{equation}
\label{eq:deltap}
\begin{split}
\Delta^\prime &= V/A = (4/3)\pi r^3/(\pi r^2) = 4r/3 \\
&= 11.0^{\prime\prime},
\end{split}
\end{equation}
where $r = \sqrt{(\text{FWHM}_a  \times \text{FWHM}_b)}$ = 8.2\arcsec. With the parameters outlined above, we plotted the derived W75N polarization angular dispersion data and fittings in Figure~\ref{fig:ADF}. We fitted the data points between 4\arcsec $<\ell <$ 8\arcsec, as scales below $\ell <$ 4\arcsec, were smaller than our synthesized beam. We set the upper fitting boundary at $\ell \approx$ 8\arcsec, as Equation~\ref{eq:adf} is valid when $\ell$ is less than a few times the beamsize ($W_1$) \citep{2009ApJ...706.1504H}. Our fitting results yielded the turbulent-to-total magnetic energy ratio, $\left<B_t^2\right>/\left<B_0^2\right>$ = 2.1 $\pm$ 0.7, and $\delta$ = 1.7 $\pm$ 0.2\arcsec. The large scale magnetic field strength was estimated as \citep{2009ApJ...706.1504H}:

\begin{equation}
\label{eq:hh09}
\begin{split}
B_0 &= \sqrt{\mu_0 \rho}\frac{\delta\nu}{\delta\theta} = \sqrt{\mu_0 \rho} \delta\nu \left[\frac{\left<B_t^2\right>}{\left<B_0^2\right>}\right]^{-1/2} \\
&= 0.8\pm0.1\,\, \text{mG},
\end{split}
\end{equation}

where $\mu_0$ is the vacuum permeability, $\rho$ is the average density of the cloud, and $\delta\nu = $ 1.5 km s$^{-1}$ is the turbulent velocity dispersion in the cloud, which was estimated from the H$^{13}$CO$^+$ (4--3) line-of-sight (LOS) velocity dispersion (see Section~\ref{subsec:line_aly}). 

Previous studies have revealed that the W75N cloud is linked to the DR21 region, and both regions are in a comparable global collapse state as a result of converging flows on large scales \citep{2010A&A...520A..49S}. Magnetic field strength measurements of DR21 cores from earlier observations range from 0.4 to 2.1 mG \citep{2013ApJ...772...69G, 2017ApJ...838..121C}, which is consistent with the magnetic field strength derived in this study for W75N.

The Alfv\'{e}nic Mach number ($M_A$), sonic Mach number ($M_s$) and the ratio of thermal-to-magnetic pressures ($\beta$) of the cloud can be calculated as:
\begin{align}
\label{eq:mams}
&M_A = \sqrt{3} \delta \nu/\nu_A,\\
&M_s = \sqrt{3} \delta \nu/c_s,\\
&\beta = 2(M_A/M_s)^2 = 2(c_s/\nu_A)^2,
\end{align}
where $\delta \nu = \delta \nu_{\text{los}}$ is the one-dimensional velocity dispersion, $\nu_A = B_0/\sqrt{\mu_0\rho}$ is the Alfv\'{e}nic velocity and $c_s = \sqrt{\gamma k_B T/(\mu m_\text{H})}$ is the sound speed at temperature $T$ using the adiabatic index $\gamma$ = 5/3 and the mean molecular weight $\mu$ = 2.33. With the average cloud temperature of 63 K, we calculated $c_s =$ 0.61 km s$^{-1}$. $\nu_A$ is calculated to be 1.0 km s$^{-1}$ and the corresponding $\beta$ value is 0.7. The calculated $M_A$, $M_s$ and $\beta$ values for the cloud are listed in Table~\ref{tab:viral_para}.

\subsection{Polarization--intensity gradient analysis}
\label{subsec:pigrad}
\begin{figure*}[ht!]
\epsscale{2}
\plotone{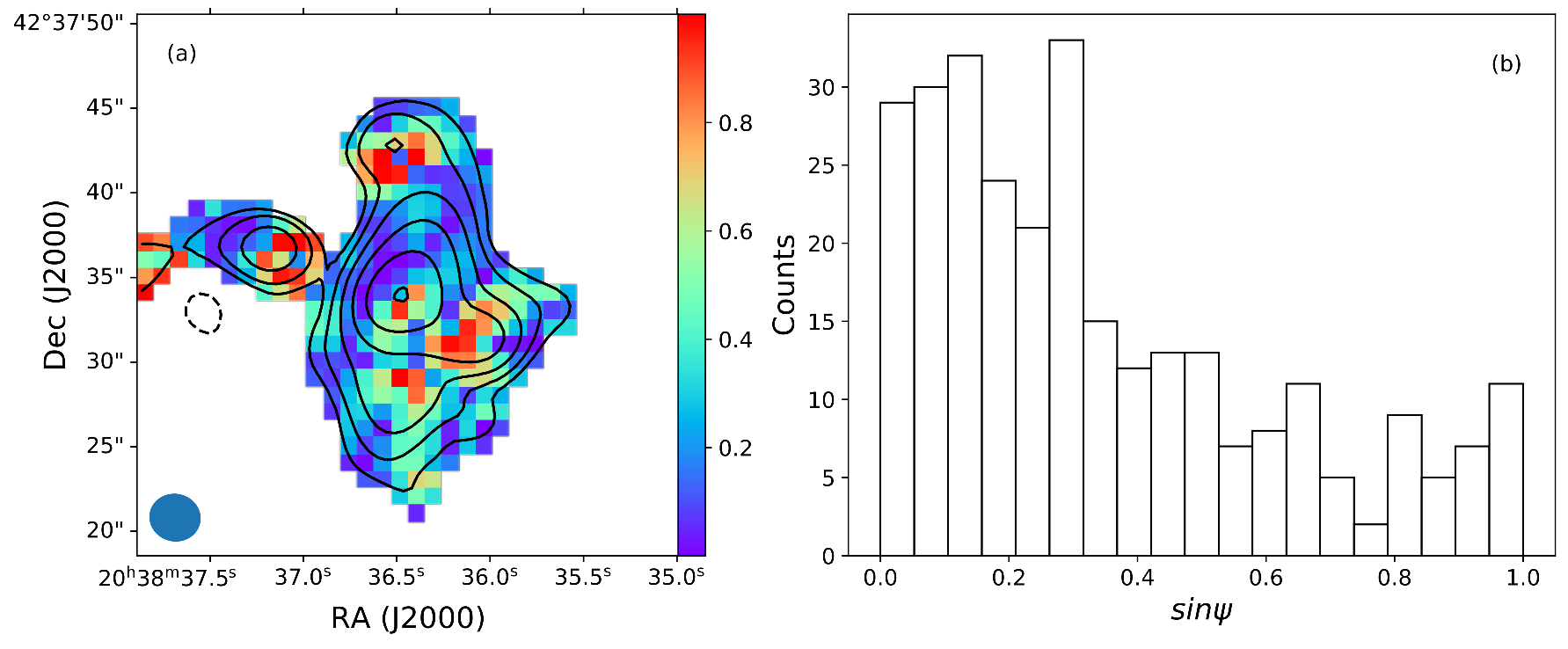}
\caption{Relevant angles for magnetic field strength derivation. (a): sin$\,\psi$--map, where $\psi$ is the difference between the orientations of the intensity gradient and local gravity. The black curves are the intensity contours as shown in Figure~\ref{fig:i}. (b): The histogram of sin$\,\psi$.}  
\label{fig:psi}
\end{figure*}

Within the framework of ideal magnetohydrodynamics (MHD), and assuming that the intensity gradient traces the direction of gas motion in the MHD force equation,  \cite{2012ApJ...747...79K} developed a technique to connect the position angle between polarization and intensity gradient orientations to the total magnetic field strength. Using this technique, we calculated the angular differences between the intensity gradient, the local gravity, and the magnetic field orientation. Figure~\ref{fig:psi} (a) displays the sin$\,\psi$--map for pixels with a detection higher than 3$\sigma$, where $\psi$ represents the difference between the intensity gradient and local gravity orientations. Assuming that mass is proportional to the detected dust emission intensity, for an intensity map with $n$ positions, the local gravity at a given position $\mathbf{r}_i$ can be calculated using the following formula \citep{2012ApJ...747...79K}:

\begin{equation}
\label{eq:lg}
\mathbf{g}(\mathbf{r}_i) \propto \sum_{j=1}^{n}\frac{I_j}{\left|\mathbf{r}_i-\mathbf{r}_j\right|^2}\cdot \mathbf{e}_{ji}, \text{ \,\,\,\,\,\,(for $j \neq i$)}
\end{equation}
where $\mathbf{e}_{ji}$ is the unit directional vector between position $\mathbf{r}_j$ and $\mathbf{r}_i$, and $I_j$ is the continuum intensity at position $\mathbf{r}_j$. Figure~\ref{fig:psi} (b) shows that the majority of sin$\,\psi$ values are small, less than 0.4, indicating that changes in the local intensity structure closely follow the local gravity. Positions with high sin$\,\psi$ values are mostly situated between intensity peaks, where the local gravity is canceled out in a particular orientation.

\begin{figure*}[ht!]
\epsscale{2}
\plotone{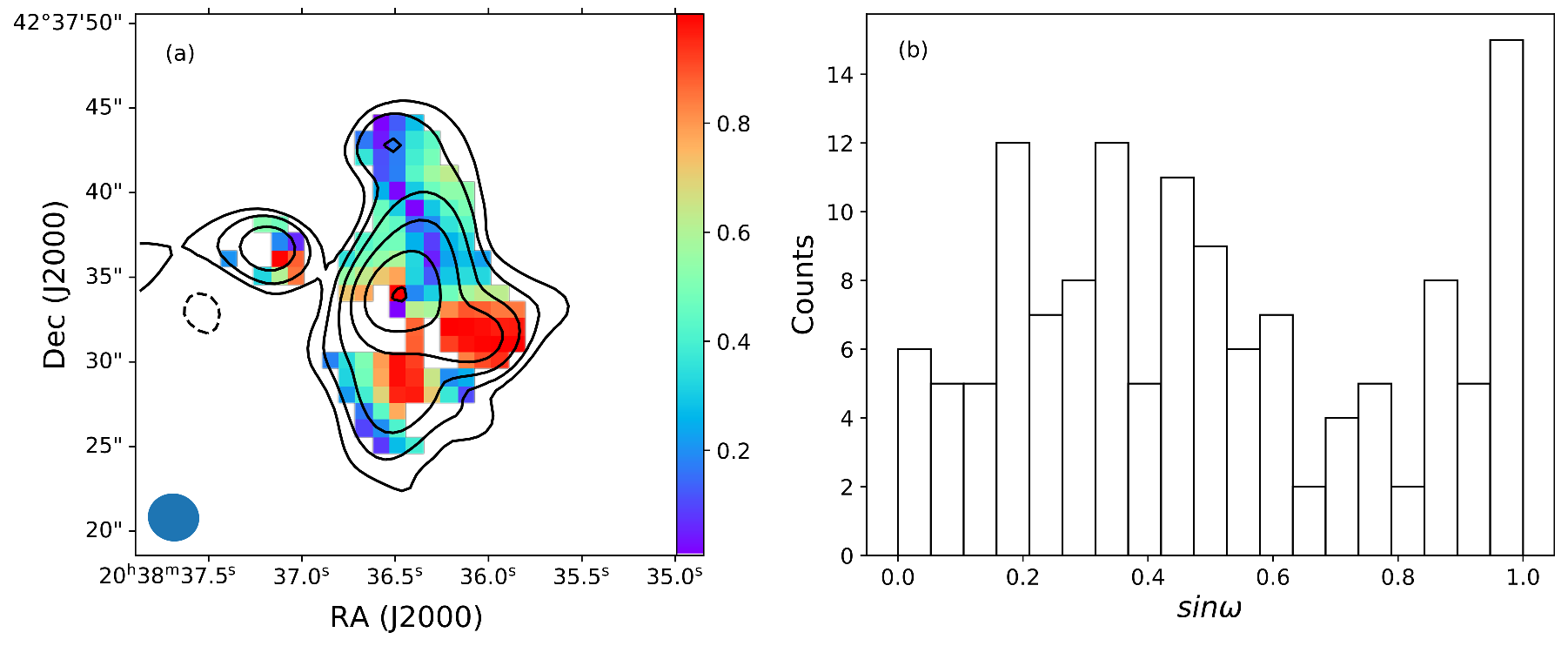}
\caption{Relevant angles for magnetic field strength derivation. (a): sin$\,\omega$--map, where $\omega$ is the difference between the magnetic field and local gravity orientations. The black curves are the intensity contours as shown in Figure~\ref{fig:i}. (b): The histogram of sin$\,\omega$.}
\label{fig:omega}
\end{figure*}

The sin$\,\omega$--map, which displays the difference between the magnetic field and local gravity orientations, is presented in Figure~\ref{fig:omega} (a), and its corresponding histogram is shown in Figure~\ref{fig:omega} (b). The sin$\,\omega$ distribution is characterized by two major peaks, one ranging from 0.2 to 0.5, and the other from 0.8 to 1.0. Regions with low sin$\,\omega$ values, particularly along the MM1 to MM[N] direction, indicate a strong alignment between the magnetic field and local gravity, resulting in a magnetic field morphology that is primarily shaped by gravity. Conversely, regions with high sin$\,\omega$ values, such as those located around the MM2, MM3, and MM4 peaks, suggest that the magnetic field is more dominant.

\begin{figure*}[ht!]
\epsscale{2}
\plotone{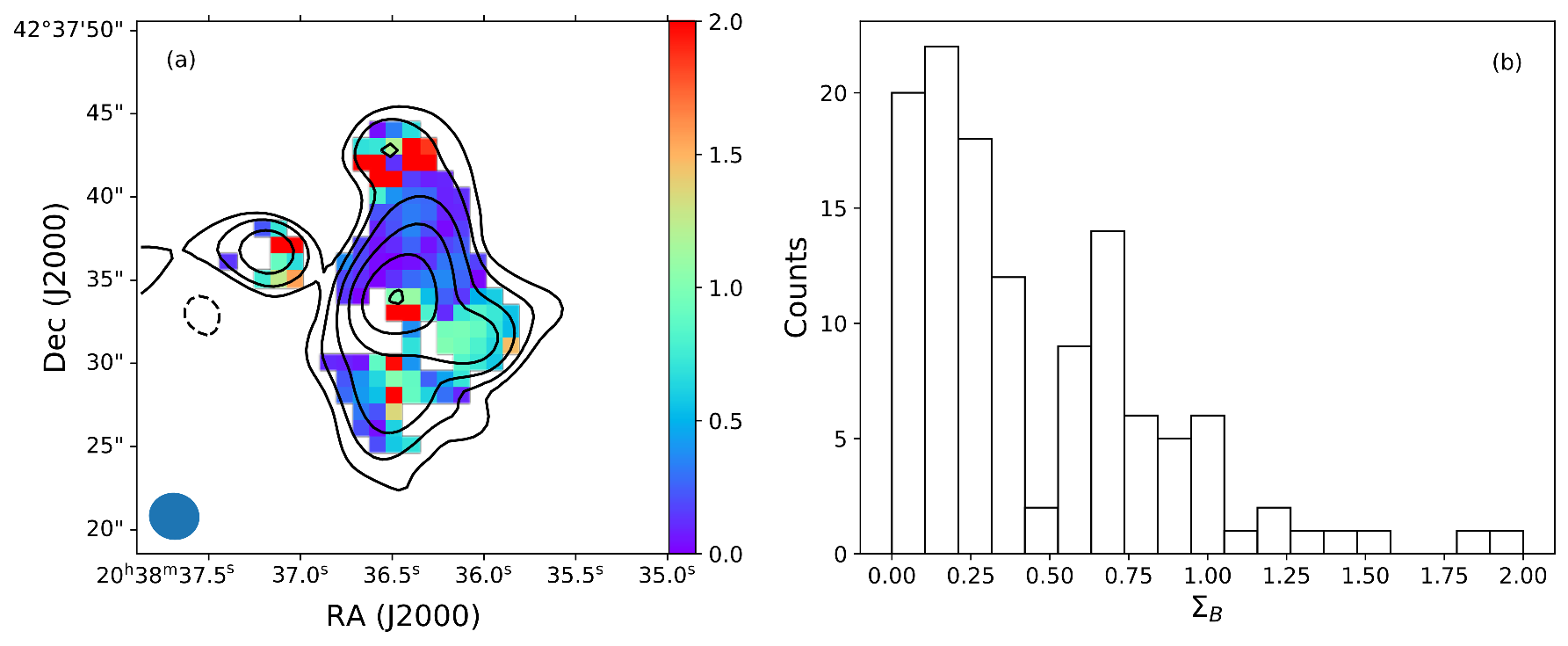}
\caption{Relevant angles for magnetic field strength derivation. (a) shows the map of field significance $\Sigma_B$. The black curves are the intensity contours as shown in Figure~\ref{fig:i}. (b): The histogram of $\Sigma_B$.}
\label{fig:SigmaB}
\end{figure*}

We also studied the magnetic field  magnetic field significance ($\Sigma_B$) to evaluate the relative importance of the magnetic field ($F_B$) in comparison to gravity ($F_G$) and pressure gradient ($F_P$) at various locations within the cloud. $\Sigma_B$ is calculated using the equation:
\begin{equation}
\label{eq:sigmaB}
\Sigma_B = \frac{F_B}{\left|F_G+F_P\right|} = \frac{\text{sin}\,\psi}{\text{cos}\,\delta}, 
\end{equation}
where $\delta$ represents the difference between the magnetic field and intensity gradient orientations. The resulting $\Sigma_B$--map and distribution are depicted in Figure~\ref{fig:SigmaB}.

Based on the results presented in Figure~\ref{fig:polvec} and Figure~\ref{fig:omega}, it appears that the MM[N] region is strongly influenced by the gravity of the main MM1 core. This gravity exerts a strong pull on the magnetic field, directing it towards the center of the cloud. Near the MM[N] peak, there is a notable discrepancy between the magnetic field and intensity gradient orientations. We conclude that the MM[N] core is a low-mass structure that is dominated by the gravity of the nearby high-mass core (MM1), similar to the case of ``Region IV" in \cite{2012ApJ...747...79K}. In such scenarios, the basic assumption that the intensity gradient traces the gas motion direction does not hold strictly, leading to high uncertainties. The $\Sigma_B$ values are dominated by large changes in $\psi$ when linked to the gravitational center of the main core. Given the lack of a clear identification of a local gravity center, the calculated $\psi$ values may be much smaller, resulting in overestimated $\Sigma_B$ values in the region, which are shown in Figure~\ref{fig:SigmaB}. 

 If we ignore the $\Sigma_B$ values near the MM[N] region, the majority of the $\Sigma_B$ values are below 1.0, particularly in the northern MM1 region, indicating that the cloud is experiencing global collapse, with the magnetic field being unable to balance the gravitational and pressure forces. Conversely, in the MM3 and MM4 core regions, the $\Sigma_B$ values are higher, suggesting that the magnetic field may be more dominant. Around the MM2 peak, the value is approximately 1, indicating that the magnetic force is comparable to the other forces.

\subsection{Molecular line emissions analysis}
\label{subsec:line_aly}
\begin{figure}[ht!]
\epsscale{1}
\plotone{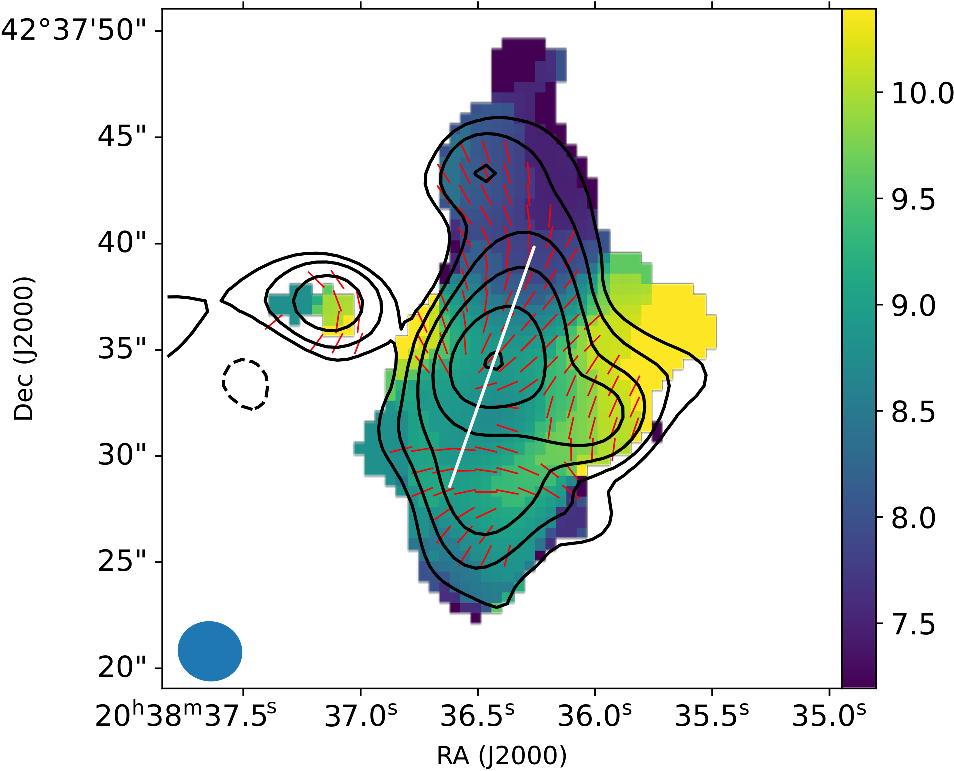}
\caption{Contour map of the dust emission superposed on the color image of the flux weighted velocity map (moment 1) of the H$^{13}$CO$^+$ (4--3) line. The color scales is in units of km s$^{-1}$. Red segments indicate the directions of the magnetic field. The path in white color show the slice position on MM1 for position--velocity analysis.}
\label{fig:H13CO_mom1}
\end{figure}

The kinematic information on the gas dynamics in the star-forming clouds enables us to probe the star formation scenario. Utilizing data from the H$^{13}$CO$^+$ (4--3) line emission, which is optically thin and devoid of self-absorption features, enables us to estimate the physical parameters of the dense cores in W75N. Figure~\ref{fig:H13CO_mom1} illustrates the first-moment map of the H$^{13}$CO$^+$ (4--3) line emission in color scale overlay on the continuum contours. The magnetic field orientations are depicted by red segments. The figure shows the contamination of high velocity components by the outflows, indicated by the redshifted lobes in Figure~\ref{fig:CO}, located to the east and west of the center MM1 region and around MM3. A significant velocity gradient from the MM[N] region to the MM1 core is observed. Based on our analysis in Section~\ref{subsec:pigrad}, the W75N cloud is undergoing global collapsing. The observed velocity gradient may be caused by gas flow from MM[N] to MM1 or cloud rotation.

\begin{figure}[ht!]
\epsscale{1}
\plotone{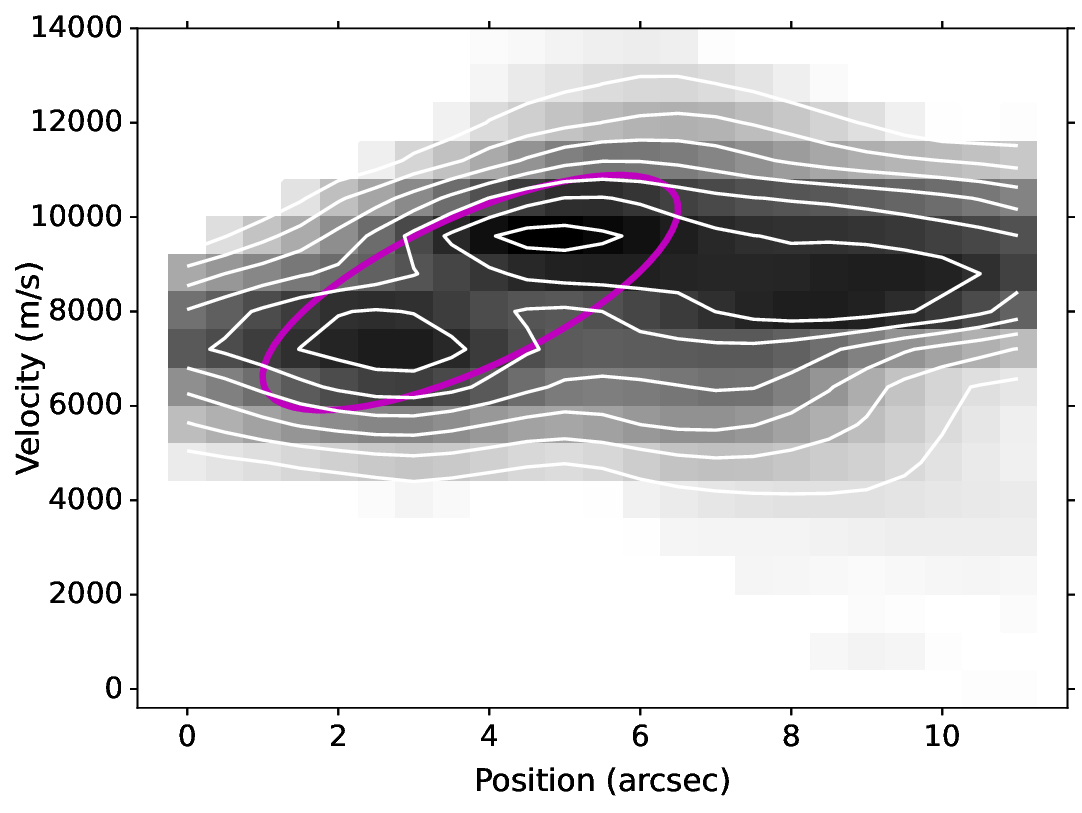}
\caption{Position–velocity map of the center region from the slice path shown in Figure~\ref{fig:H13CO_mom1}. The purple ellipse shows the best fit to velocity gradient.}
\label{fig:H13CO_PV_MM1}
\end{figure}

To avoid the contamination from the outflows, we perform the position--velocity (PV) analysis to model the velocity gradient along the vertical white color path (PA = --20\deg) shown in Figure~\ref{fig:H13CO_mom1}. The PA angle is chosen to be perpendicular to the large-scale outflow shown in Figure~\ref{fig:CO} and consistent with the disk-like structure unidentified by \cite{2002MNRAS.330..349H} and \cite{2021A&A...655A..86V}. The ellipse resulting from the best 2D Gaussian fit, represented in Figure~\ref{fig:H13CO_PV_MM1}, indicates a slope angle of 24\deg\, and a calculated velocity gradient of approximately 0.9 km s$^{-1}$ arcsec$^{-1}$. If the observed gradient is due to cloud rotation, it corresponds to a rotation velocity of $\omega$ = 1.4$\times$10$^{-4}$ yr$^{-1}$, resulting in $(\omega/B)_{\text{obs}}$ = 1.7$\times$10$^{-7}$ yr$^{-1}\mu$G$^{-1}$. Depending on the magnetic field strength and rotation velocity, the evolution of a collapsing dense core can be regulated either by centrifugal forces or magnetic forces. We define a centrifugal critical parameter $\chi$, which is the ratio of the observed $(\omega/B)_{\text{obs}}$ to the critical $(\omega/B)_{\text{crit}}$ \citep{2005MNRAS.362..369M}: 

\begin{equation}
\label{eq:Mchi}
\begin{split}
\chi &= \frac{(\omega/B)_{\text{obs}}}{(\omega/B)_{\text{crit}}} \\
&= \frac{(\omega/B)_{\text{obs}}}{1.69\times 10^{-7} (c_s/0.19 \text{ km s}^{-1})^{-1} \text{yr}^{-1} \mu \text{G}^{-1}}.
\end{split}
\end{equation}
Given $c_s =$ 0.61 km s$^{-1}$, $(\omega/B)_{\text{crit}}$ is calculated to be 5.3 $\times$10$^{-8}$ yr$^{-1}\mu$G$^{-1}$ and $\chi$ value for the cloud is 3.5, which is greater than 1. The centrifugal forces dominates the dynamics of the collapse over the magnetic field.

We determined the turbulent velocity dispersion, $\delta \nu_{\text{los}}$, by fitting the line width of the H$^{13}$CO$^+$ (4--3) spectrum. Since the molecular weight is high and $T_d$ (temperature) is low, the impact of thermal velocity dispersion is negligible. To eliminate the contribution of large-scale velocity motion within the cloud, we applied a method that shifts the velocity of a spectrum for each spatial pixel by the centroid velocity indicated in the moment 1 map (refer to Figure~\ref{fig:H13CO_mom1}) to remove the large-scale velocity field. This technique shifts the average velocity of each pixel to zero, isolating only the turbulent component. The turbulent velocity is then determined by fitting a Gaussian profile to the intensity--velocity curve. The final estimated value for $\delta \nu_{\text{los}}$ is approximately 1.5 km s$^{-1}$.

The ratio of the turbulent to magnetic energy $\beta_{\text{turb}}$ is usually calculated using the Alfv\'{e}nic Mach number:
\begin{equation}
\label{eq:beta_t}
\beta_{\text{turb}} = M_A^2 = 3(\delta\nu_{\text{los}}/\nu_A)^2.
\end{equation}
The $\beta_{\text{turb}}$ for the entire cloud is calculated to be 10.9, indicating the turbulent energy dominates the magnetic energy. 

The relative importance between the magnetic field and the gravity of individual sources can be estimated by the magnetic critical parameter $\lambda$, which is the mass-to-flux ratio in units of the critical value $1/(2\pi\sqrt{G})$ \citep{1976ApJ...210..326M,1978PASJ...30..671N,2004Ap&SS.292..225C}:
\begin{equation}
\label{eq:Mphi}
\begin{split}
\lambda &= \frac{(M/\Phi_B)_{\text{obs}}}{(M/\Phi_B)_{\ ntext{crit}}} \\
&= 7.6\times 10^{-21}\frac{N_{\text{H}_2}/(\text{cm}^{-2})}{B/(\mu \text{G})}.
\end{split}
\end{equation}
The calculated $\lambda $ value for W75N is about 2.0, indicating gravity dominates the magnetic field.

Table~\ref{tab:viral_para} lists the viral parameters of W75N. The cloud has $M_s >$ 1, revealing that non-thermal motions are supersonic. The $M_A$ value is greater than 1, indicating that turbulent energy is stronger than magnetic energy. These supersonic and super-Alfvénic Mach numbers imply the presence of strong non-thermal motions in the cloud. The $\beta$ value is less than one, indicating that although weaker than the non-thermal pressure, the magnetic pressure is stronger than the thermal pressure. This $M_s > M_A > 1 > \beta$ relationship has been previously observed in other high-mass forming regions, such as the DR21 cores \citep{2017ApJ...838..121C}.

The average $\lambda$ value for the cloud is 2.0, indicating it is under going a global collapse. The estimated $B_0$ and $\lambda$ values in this work are consistent with the results ($B_0$ = 0.3 -- 1.2 mG and $\lambda$ = 0.6 -- 2.2) obtained by \cite{2021ApJ...912..159P}. The cloud exhibits a large scale velocity gradient, but it is unclear whether it is due to gas in-fall or cloud rotation. If the cloud is rotating, the high $\chi$ value suggests that the centrifugal force dominates the magnetic field force.

\begin{deluxetable}{cccccccc}
\tablecaption{Viral parameters of W75N}
\label{tab:viral_para}
\tablewidth{0pt}
\tablehead{
\colhead{$\delta \nu_{\text{los}}$/(km s$^{-1}$)} & \colhead{$M_s$} & \colhead{$M_A$} & \colhead{$\beta$} & \colhead{$\beta_{\text{turb}}$} & \colhead{$\lambda$} & \colhead{$\chi$}
}
\startdata
 1.5 & 5.6 & 3.3 & 0.7 & 10.9 & 2.0 & 3.5 \\
\enddata
\end{deluxetable} 

\section{Discussion} 
\label{sec:disc}

Observations of the Zeeman effect towards maser sources at small scales have been used to derive magnetic field strengths in the line-of-sight (LOS) direction. The magnetic field strength derived from Zeeman pairs of opposite circular polarization ranges from +8 to -8 mG using OH masers at 1665, 1667, and 1720 MHz \citep{2002MNRAS.330..349H}. \cite{2007ApJ...656..952F} detected a strong magnetic field source of about 40 mG near VLA 2. \cite{2009A&A...506..757S} observed the 6.7 GHz methanol maser using the European VLBI network and found that the Zeeman-splitting measurements indicated the LOS magnetic fields in the maser regions ranging from 11 to 16 mG. In contrast, the magnetic field strength measured from observations of the 22 GHz water masers is about 1000 mG \citep{2011A&A...527A..48S}, which is much higher than those from the methanol maser observations. Recently, \citet{surcis2023monitoring} measured --764 mG $< B^\text{VLA1}<$ --676 mG and --355 mG $< B^\text{VLA2}<$ --2426 mG in the LOS direction with 22 GHz water maser observations. These high-resolution (typically around 10$^2$ AU) maser observations detected much higher LOS magnetic field strengths at small scales in protostellar envelopes. The hydrogen number densities of those regions estimated using the empirical equation $B \propto n^{0.65}_{\text{H}_2}$ \citep{2010ApJ...725..466C} range from 10$^8$ to 10$^{10}$ cm$^{-3}$ \citep{surcis2023monitoring}. It is not straightforward to compare the results from our work using thermal dust emission to those from maser observations arising from non-thermal processes. Based on the findings from \cite{2022ApJ...925...30L}, the density and magnetic field strength (0.85$\times 10^{6}$ cm$^{-3}$ and 0.8 mG) from this work indicate that the cloud is in a magnetically supercritical phase.


\cite{1995A&A...294..831V} conducted JCMT observations towards the compact source W75N-IRS1 using a beamsize of 12\arcsec\, at 870 $\mu$m. They detected one magnetic segment with a PA = 145\deg $\pm$ 5\deg\, and estimated a magnetic field strength of $B$ = 0.8 mG using a simple statistical relation between the magnetic field strength and the gas density. Their magnetic field strength and PA are consistent with the mean field of our results. Using JCMT, \cite{1998A&A...333L..23G} obtained a similar magnetic field position angle of 153\deg $\pm$ 22\deg\, at 1100 $\mu$m, while at 450 $\mu$m, the derived magnetic field was 37\deg $\pm$ 9\deg. The change in magnetic field PA could be attributed to the twisted magnetic field lines around the region. The net magnetic field value could change as the beamsize varies. Similarly, the maser observations obtained magnetic fields perpendicular to our submillimeter polarization observations because the maser observations were at milliarcsecond (mas) resolution to trace the compact \ion{H}{2} regions. The magnetic field could twist significantly from mas to arcsec scales.

\begin{figure*}[ht!]
\epsscale{1.5}
\plotone{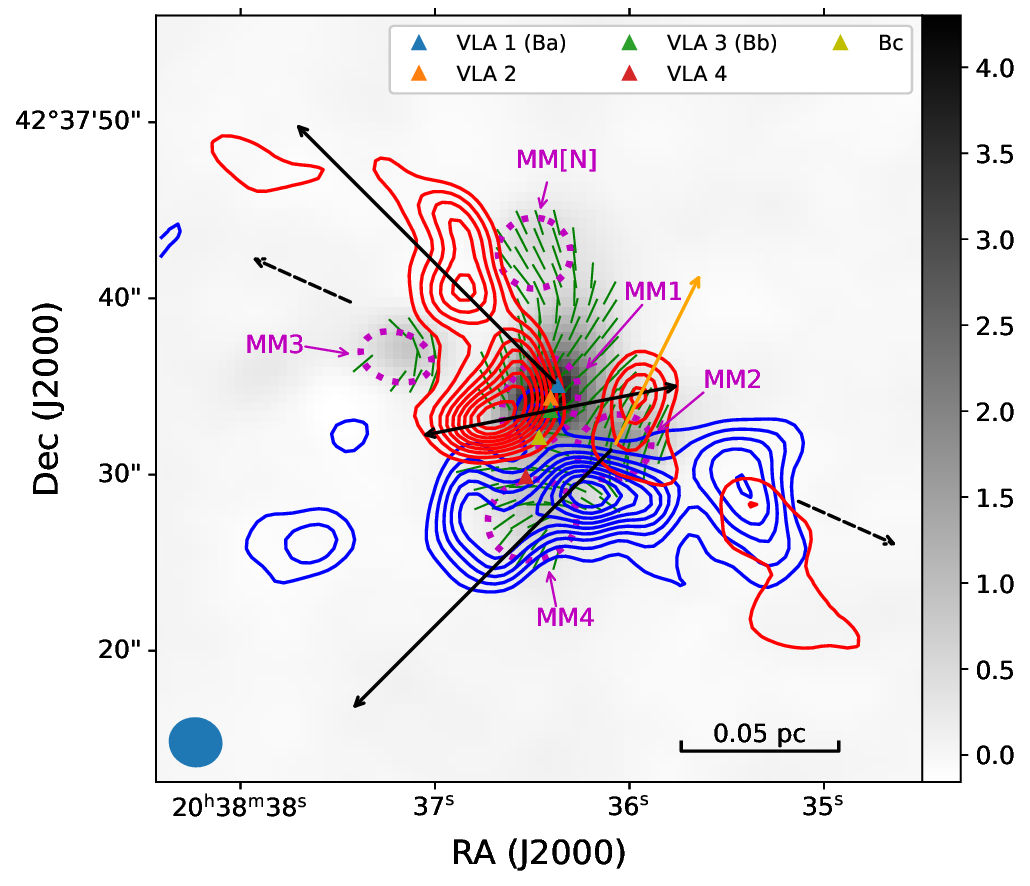}
\caption{Integrated CO (3--2) blueshifted and redshifted emission contours from -18.0 to 0 km s$^{-1}$ and 20.0 to 28.0 km s$^{-1}$, respectively. The peaks for the blueshifted and redshifted emission are 275.3 mJy~beam$^{-1}$ km s$^{-1}$ and 191.6 mJy~beam$^{-1}$ km s$^{-1}$. Contours begin at 10\% of the peak emission and continue at increments of 10\%. The continuum emission shown in background is identical to that in Figure~\ref{fig:i}. The two dashed arrows indicated the direction of the bipolar outflow from \cite{1994A&A...284..215H}. The three solid arrows are from \cite{2003ApJ...584..882S}. Compact sources are labeled as filled triangles. The green segments are the magnetic field from Figure~\ref{fig:polvec}. The dashed ellipses in magenta show the dense cores from} Figure~\ref{fig:i} (b).
\label{fig:CO}
\end{figure*}

In Figure~\ref{fig:CO}, we present the CO (3–2) blueshifted and redshifted emission contours from our work. We chose the velocity boundaries of the blueshifted (-18.0 to 0 km s$^{-1}$) and redshifted (20.0 to 28.0 km s$^{-1}$) emissions to be symmetrical with respect to the cloud's $\nu_{\text{LSR}}$  = 10.0 km s$^{-1}$, as reported by \citep{2003ApJ...584..882S}. The compact sources VLA 1 (Ba), VLA 2, VLA 3 (Bb), Bc, and VLA 4 are marked as filled triangles, and the dense cores of the cloud from Figure~\ref{fig:i} are labeled using dashed ellipses. The black dashed arrows indicate the direction of the bipolar outflow (66\deg) from \cite{1994A&A...284..215H}, and the three black solid arrows from \cite{2003ApJ...584..882S} show the outflow orientations for the redshifted component (45\deg, started from VLA 1), blueshifted component (135\deg, started from MM2), and the bipolar outflow from VLA 3 (101\deg, centered at VLA 3). \cite{1997ApJ...489..744T, 2009A&A...506..757S} also suggested that VLA 1 powers the large-scale molecular bipolar outflow of W75N(B). 

We found that the main outflows centered at VLA 1 and MM2 from \cite{2003ApJ...584..882S} match well with the high velocity gas detected in our CO (3--2) emission map. However, we did not detect the blueshifted components of the bipolar outflows from VLA 3 to the west of the source. We propose the existence of another outflow centered at MM2, extending in a direction almost opposite to that of MM4, indicated by the orange arrow in Figure~\ref{fig:CO}. The bipolar outflows originating from the MM2 core drag and align the magnetic field lines in the MM2 and MM4 regions. In addition, we found enhanced dust polarization along the cavity walls of the redshifted lobe of the outflow, specifically around the MM3 region. The magnetic field lines in the MM[N] and MM1 regions are shaped by gas infall from the MM[N] to MM1 core. These findings are consistent with the results of the polarization angle analysis presented in Section~\ref{subsec:pol}.

The overall $\lambda$ is greater than 1, and the $\Sigma_B$ values shown in Figure~\ref{fig:SigmaB} (b) predominantly fall below 1, indicating that the W75N cloud is undergoing global collapsing. In the MM2 and MM4 regions, while the $\Sigma_B$ values increase, they still remain primarily below 1, as these regions are dominated by gravity and pressure gradient. The magnetic field is also shaped by the outflows from the MM2 core. If the large scale velocity is from cloud rotation, the average cloud $M_A$ = 3.3 and $\chi = 3.5$, indicating that turbulence and the centrifugal force dominate over magnetic field.

\section{Conclusion} 
\label{sec:sum}

We present 345 GHz polarization observations of the W75N region using the SMA interferometer. We estimated the physical parameters of the dense structures in the region from the dust continuum emission. Our analysis reveals a uniform distribution of polarization angles within each dense structure. We used the ADF method to study the POS magnetic field and estimated a large-scale magnetic field component of 0.8 $\pm$ 0.1 mG. We also investigated the dynamical state of the cloud by analyzing the polarization-intensity gradient and the H$^{13}$CO$^+$ (4--3) line data. Our findings suggest that the W75N region is undergoing global collapsing due to the weaker magnetic field force compared to other forces. We observed that the magnetic field around the MM[N] and MM1 regions is aligned by gas infall, while in the MM2 and MM4 regions, the magnetic field is shaped by outflows from the MM2 core. We also observed enhanced dust polarization along the cavity walls around the MM3 region.

\begin{acknowledgments}
This work was partially supported by the program Unidad de Excelencia María de Maeztu CEX 2020-001058-M. JG also acknowledges support by the grant PID 2020-117710 GB-I00 (MCI-AEI-FEDER,UE).
\end{acknowledgments}

%

\bibliography{w75n}{}
\bibliographystyle{aasjournal}



\end{document}